\documentclass[preprint,preprintnumbers,amsmath,amssymb]{revtex4}
\usepackage{amsfonts}    % list packages between braces
\usepackage{amssymb}
\usepackage{latexsym}
\usepackage{eepic}
\usepackage{graphicx}
\usepackage[usenames,dvipsnames]{color}

\newcommand{\al}{\alpha}
\newcommand{\pa}{\partial}

\newcommand{\ta}{\tau}

\newcommand{\Om}{\Omega}
\newcommand{\om}{\omega}

\newcommand{\De}{\Delta}

\newcommand{\rar}{\rightarrow}

\begin{document}

\title{Particular Integrability and (Quasi)-exact-solvability}

\author{A.V.~Turbiner}
\email{turbiner@nucleares.unam.mx}
\affiliation{Instituto de Ciencias Nucleares, Universidad Nacional
Aut\'onoma de M\'exico, Apartado Postal 70-543, 04510 M\'exico,
D.F., Mexico}

\date{August 6, 2012}

\begin{abstract}
A notion of a particular integrability is introduced when two operators commute on a subspace of the space where they act. Particular integrals for one-dimensional (quasi)-exactly-solvable Schroedinger operators and Calogero-Sutherland Hamiltonians
for all roots are found.
In the classical case some special trajectories for which the corresponding particular constants of motion appear are indicated.
\end{abstract}

%\pacs{31.15.Pf,31.10.+z,32.60.+i,97.10.Ld}

\maketitle

Assume that $\varphi_0$ is a common eigenfunction for two algebraically-independent operators $A$ and $B$:\ $A\varphi_0=a_0\varphi_0$ and $B\varphi_0=b_0\varphi_0$.
It is evident that the commutator $[A, B]\varphi_0 = 0$.
The best-known example is related to the harmonic oscillator Hamiltonian in the second quantization representation,
$H (\hat p, x) = \{ a^+ a^-\}_{+}  \equiv A$, then $B (\hat p, x) = a^-$ commutes with $H$ over
the vacuum, $[H, a^-]|0>=0$.
By taking a straightforward classical analogue $H (\hat p, x) \rar H (p, x)$ and
$B (\hat p, x) \rar B (p, x)$, one can ask whether there exists a trajectory for which
the Poisson bracket $\{ H, B \}$ vanishes? This trajectory really exists and corresponds
to $p=x=0$, when the particle stands at the bottom of the potential. For this trajectory
the energy $E$ is equal to zero as well as $B=0$. It is not periodic unlike all other
trajectories at $E \neq 0$. Such a trajectory we will call {\it special}.

It is evident that if for operators $A$ and $B$ there exists a number of common eigenfunctions $\{\varphi_0\}$, the commutator $[A, B]$ annihilates a space
$V_p$ spanned by $\{\varphi_0\}$. At the same time $V_p$ is the invariant subspace for
both $A$ and $B$. In more general terms, if we take two operators $A, B$ acting
on a space $V$ and assume that for a subspace $V_p \subset V$, the commutator $[A, B]: V_p \rar \{0\}$ and both operators $A, B$ have $V_p$ as an invariant subspace $A, B: V_p \rar V_p$, then the operators $A, B$ are called {\it particular} integrals ($\pi$-integrals).
It can easily be shown if $[A, B]: V_p \rar \{0\}$, and the operator $A$ has $V_p$ as an invariant subspace and non-degenerate, with all eigenvalues of multiplicity one, then the operator $B$ has $V_p$ as an invariant subspace.
Thus, $\pi$-integrability implies the existence of the common eigenfunctions of
two spectral problems $A(B) \varphi = a(b) \varphi$, where  $a(b)$ are spectral
parameters. If the dimension of $V_p$ is finite, $\dim V_p < \infty$, the operators
become {\it quasi-exactly-solvable}, for which a finite number of eigenfunctions can be found explicitly by algebraic means (for a review, see \cite{Turbiner:1994}). It allows us to connect quasi-exact-solvability with particular integrability. It is worth noting an important particular case of $\pi$-integrals when $A: V_p \rar V_p$ and $B: V_p \rar \{0\}$, hence $V_p$ is the kernel space or the space of zero modes of $B$. Another example of $\pi$-integrals is when $A(B)$ are operators which depend on different variables; the bispectral problem by Duistermaat-Gr\"{u}nbaum \cite{DG:1986} is a particular case of it.
If the common eigenfunctions form a complete basis, hence, $V_p = V$,
the commutator $[A, B]=0$. This situation corresponds to a standard integrability.

The classical counterpart of $\pi$-integrability implies that the Poisson bracket
$\{ A, B \}$ may vanish on some {\it special} trajectories. These trajectories can be called superintegrable: they are characterized by a larger number of constants of motion
than generic ones.
It is not clear whether quantum  $\pi$-integrability implies always the existence of physical special trajectories; they might be complex or might not exist at all. The goal of this paper is to study quantum $\pi$-integrals of the one-dimensional (quasi)-exactly-solvable Schroedinger equations, the Calogero-Sutherland systems for all roots and their classical counterparts, the classical $\pi$-integrals and related
special trajectories.

Take the algebra $gl_2$ in $(n+1)$-dimensional representation realized by the
first order differential operators
\[
J^-\ =\ \frac{d}{dt} \ ,
\]
\begin{equation}
\label{sl2}
 J^0_n\ =\ t \frac{d}{dt} - n \  ,\ T^0\ =\ 1
\end{equation}
\[
J^+_n\ =\ t^2 \frac{d}{dt} - n t\ =\ t J^0_n\ ,
\]
where $n=0,1,\ldots$, $T^0$ is the center element. The finite-dimensional representation space is the space of polynomials of degree not higher than $n$,
\begin{equation}
\label{pn}
  {\cal P}_{n} \ = \ \langle 1, t, t^2, \dots , t^n
\rangle \equiv (t^k|\ 0\leq k \leq n)\ .
\end{equation}
The generator $J^-$ is lowering operator or filtration (of the grading -1)
\[
  J^- : {\cal P}_{n} \mapsto {\cal P}_{n-1}\ .
\]
Hence, ${\cal P}_{n}$ is the kernel of $(J^-)^{n+1}$. It is evident that any non-trivial element $h$ of the universal enveloping algebra $U_{gl_2}$ taken in the realization (\ref{sl2}) commutes with
\begin{equation}
\label{ino}
i_n^{(-1)} \equiv (J^-)^{n+1}\ ,
\end{equation}
as well as with
\begin{equation}
\label{ink}
i_n^{(k)} \equiv (J^-)^{n-k} \prod_{j=0}^k (J^0_n + j)\ ,\ k=0,1,\ldots n\ ,
\end{equation}
over ${\cal P}_{n}$, namely,
\begin{equation}
\label{hi}
  [h, i_n^{(k)}]: {\cal P}_{n} \ \mapsto \ 0\ ,\ k=-1,0,1,\ldots n\ .
\end{equation}
Hence, with respect to $h$ any $i_n^{(k)}, k=-1,0,1,\ldots n$ is $\pi$-integral over ${\cal P}_{n}$ and its grading $\deg(i_n^{(k)})=-(n-k)$. For any $i_n^{(k)}$ the space ${\cal P}_{n}$ is the space of zero modes. Hence, all these $\pi$-integrals are in involution,
\[
[i_n^{(k)}, i_n^{(m)}]: {\cal P}_{n} \ \mapsto \ 0\ .
\]
Operators $i_n^{(m)},\ m=-1,0,1,\ldots n$ are generating elements of some infinite-dimensional algebra. This is the algebra of differential operators for which any element has the space ${\cal P}_{n}$ as the kernel.

Any element $e$ of the universal enveloping algebra $U_{gl_2}$, that does not contain the generator $J^+_n$, preserves the infinite flag ${\cal P}$, 
\[
{\cal P}_0 \subset  {\cal P}_1 \subset {\cal P}_2 \subset \ldots
 \subset  {\cal P}_p  \subset \ldots \ ,
\]
therefore
\begin{equation}
\label{ei}
  [e, i_m^{(k)}]: {\cal P}_{p} \ \mapsto \ 0\ ,\ k=-1,0,1,2,\ldots m
\end{equation}
for any $m=0,1,\ldots$ and $p\leq m$. It seems evident, that there is no $\pi-$integral in a form of an element of $U_{gl_2}$ other than a polynomial $f(e)$ which commutes with $e$ over any ${\cal P}_{p}$.

Take a quadratic element $h_2$ of the universal enveloping algebra $U_{gl_2}$ in the realization (\ref{sl2}),
\[
h_2 =  c_{++} J^+_n J^+_n  + c_{+0} J^+_n  J^0_n  + c_{+-} J^+_n  J^-  +
c_{0-} J^0_n  J^-  + c_{--} J^-  J^-  +
\]
\begin{equation}
\label{qes}
 c_+ J^+_n  + c_0 J^0_n  + c_- J^-  + c\ ,
\end{equation}
where $c_{\al \beta}, c_{\al}, c \in {\mathbf R}$ are parameters, their number is
$par (h_2) = 9$. Substituting (\ref{sl2}) into (\ref{qes}) we obtain the Heun operator
\begin{equation}
\label{qesX}
h_2(t)\ =\ - P_{4}(t) \frac{d^{2}}{dt^{2}} \ +\ P_{3}(t) \frac{d}{dt} \
+\ P_{2}(t) \ ,
\end{equation}
where the $P_{j}(t)$ is a polynomial of $j$th order with coefficients
related to  $ c_{\al \beta}, c_{\al}$ and $n$, that can be easily calculated.
In general, the operator (\ref{qesX}) is diagonalizable, it has $(n+1)$ polynomial eigenfunctions in the form of a polynomial in $t$ of $n$th degree, hence, this is a quasi-exactly-solvable operator.
Every polynomial eigenfunction is an element of ${\cal P}_{n}$. Moreover, it is known that the operator $h_2(t)$ is the most general quasi-exactly-solvable second-order differential operator acting in the space of monomials \cite{Turbiner:1992}. In explicit form,  the $\pi$-integral $i_n^{(-1)}$ for $h_2(t)$ has the form,
\begin{equation}
\label{in0X}
    i_n^{(-1)}(t)\ =\ \frac{d^{n+1}}{dt^{n+1}}\ ,
\end{equation}
while the $\pi$-integral $i_n^{(k)}$ is
\begin{equation}
\label{inkX}
    i_n^{(k)}(t)\ =\ \frac{d^{n-k}}{dt^{n-k}} \prod_{i=0}^k (t \frac{d}{dt} - n + i)\ ,\quad k=0, 1,\ldots n\ .
\end{equation}
Hence, $h_2(t)$ and all $i_n^{(k)}(t),\ k=-1, 0, 1,\ldots n$ have $(n+1)$ common eigenfunctions in the form of a polynomial in $t$ of $n$th degree. For any $i_n^{(k)}$ these polynomial eigenfunctions are the zero modes.
%It can be shown that the operator (\ref{qesX}) is the most general second order differential operator which commute with (\ref{in0X})-(\ref{inkX}) over the space ${\cal P}_{n}$.
This explains quasi-exact-solvability of (\ref{qesX}) as a consequence of the existence of the $\pi$-integral $i_n^{(k)}, k=-1,0,1,\ldots n$.

The commutativity (\ref{hi}) remains unchanged under a gauge (similarity) transformation of operators and space ${\cal P}_{n}$ with consequential change of variable $t \mapsto x$,
\begin{equation}
\label{hiZ}
  [{\cal H}(x), {\cal I}_n^{(k)}(x)]: {\cal V}_{n} \ \mapsto \ 0\ ,
\end{equation}
where
\[
    {\cal H}(x)\ =\ e^{ -A(t)}\ h(t)\ e^{ A(t)}|_{t=t(x)}\ ,\
    {\cal I}_n^{(k)}(x)\ =\ e^{ -A(t)}\ i_n^{(k)}(t)\ e^{ A(t)}|_{t=t(x)}
 \]
and
\begin{equation}
\label{vn}
  {\cal V}_{n} \ = \  e^{-A(t(x))}\ <t^k(x)|\ 0 \leq k \leq n>\ .
\end{equation}
Taking the gauge phase
\[
A(t) = \int \bigg(\frac{P_3}{P_4}\bigg)dt - log x'\ ,
\]
and changing the variable
\[
\ x(t) = \pm\int \frac{dt}{\sqrt{P_4}}\ ,
\]
we arrive at ${\cal H}(x)$ in the form of the Schroedinger operator
\begin{equation}
\label{Hz}
 {\cal H}(x) = - \frac{d^2}{dx^2} + V(x)\ ,\
 V(x) = (A'_t)^2 (t(x)) - A''_{tt} (t(x)) + P_2 (t(x)) \ .
\end{equation}
Taking the set of the coefficients $c_{\al \beta}, c_{\al}$ in (\ref{qes}) accordingly,  one can obtain the Hamiltonians of all ten known $1D$ quasi-exactly-solvable Schroedinger equations with hidden algebra $gl_2$ \cite{Turbiner:1988, Turbiner:1989, Kamran:1993}. In this case ${\cal V}_{n}$ is the subspace of the Hilbert space. The gauge transformed $\pi$-integrals (\ref{in0X}), (\ref{inkX}) become quantum $\pi$-integrals,
\begin{equation}
\label{Inz}
 {\cal I}_n^{(-1)}(x) = (\frac{d}{dt} + A'_t)^{n+1}|_{t=t(x)} \ ,\ {\cal I}_n^{(k)}(x) = (\frac{d}{dt} + A'_t)^{n-k}\prod_{j=0}^k (t \frac{d}{dt} + t A'_t - n + j)|_{t=t(x)} \ .
\end{equation}
Replacing the momentum operator by classical momentum
\begin{equation}
\label{quant}
\frac{d}{dt} = i \hat{p} \rar i p
\end{equation}
in (\ref{Inz}), the classical counterparts of $\pi$-integrals occur,
$I_n^{(k)}(p, x) \equiv {\cal I}_n^{(k)}(p, x)$. In general, they have a form
of a finite-degree polynomial in $p$ with singular coefficient functions, hence, they are
singular functions on the phase space and can cease to exist some trajectories.
Usually, the only classical $\pi$-integral function, which is non-singular, is $I_n^{(n)}(p, x) \equiv I_n(p, x)$. A natural question to ask is whether there exists a special trajectory(ies) $\{s(p,x)\}$ for which $I_n(p, x)$ is a constant. For those trajectories the Poisson bracket vanishes, $\{ H(p,x), I_n(p,x) \}_{s(p,x)} = 0$. For all known quasi-exactly-solvable Hamiltonians, special trajectories exist. They correspond to a particle at rest at extreme of the potential.

Let us consider a particular case of QES sextic anharmonic oscillator, which is the only $1D$ polynomial potential with QES property. It is described by a quadratic combination in $gl_2-$generators (1),
\begin{equation}
\label{qes-p}
 h_{2,p} = -4 J^0_n J^-_n + 4a J^+_n + 4b J^0_n - 2(n+1+2q) J^-_n + 2bn
\end{equation}
or, as the second order differential operator,
\[
h_{2,p}(t)= -4t \frac{d^2}{dt^2} + 2(2at^2+2bt-1-2q) \frac{d}{dt} - 4ant  \ ,
\]
where $t \in [0, \infty)$ and $a>0, \forall b$ or $a \geq 0, b>0$ . Putting
$t=x^2$ and choosing the gauge phase as follows $$A=\frac{a t^2}{4} + \frac{bt}{2} -
\frac{q}{2} \ln{t}\ ,$$ we arrive at the Hamiltonian (\ref{Hz}) with the
potential \cite{Turbiner:1988} (see also \cite{Turbiner:1987})
\begin{equation}
\label{qes-p-V}
 V_6(x) = a^2 x^6 + 2 a b x^4 + [b^2 - (4n+3+2q)a] x^2 - b(1+2q),
\end{equation}
for which $q=0\ (q=1)$ and the first $(n+1)$ eigenfunctions, even
(odd) in $x$, can be found algebraically. Of course, the number of
those ``algebraic" eigenfunctions is nothing but the dimension
of the irreducible representation of the $gl_2-$algebra (\ref{sl2}).
These $(n+1)$ ``algebraic" eigenfunctions have the form
\begin{equation}
\label{psin}
 \Psi^{(n)} (x) \ =\ x^q p_n (x^2)  e ^ {-\frac{a x^4}{4} - \frac{b x^2}{2}}\ ,
\end{equation}
where $p_n$ is a polynomial of degree $n$.
Quantum $\pi$-integral ${\cal I}_n (x) \equiv {\cal I}_n^{(n)}(x)$ becomes
\begin{equation}
\label{Inz6}
 {\cal I}_n(x)\ =\ \frac{1}{2^{n+1}} \prod_{j=0}^n (x \frac{d}{dx} + {a x^4} + {b x^2} -
                    q - 2n + 2j) \ .
\end{equation}
It is easy to give a direct proof that the Hamiltonian ${\cal H}_6(x)$ with the potential (\ref{qes-p-V}) commutes with ${\cal I}_n(x)$,
\[
     [- \frac{d^2}{dx^2} + V_6(x),\ {\cal I}_n(x)]: {\cal V}_{n}\ \mapsto \ 0\ .
\]
Replacing the operator of momentum by the classical momentum (\ref{quant}) in ${\cal H}_6(x)$ and ${\cal I}_n(x)$ we get a classical Hamiltonian $H_6$ and a classical analogue of the $\pi-$integral,
\begin{equation}
\label{Inz6c}
 {I}_n(p,x)\ =\ \frac{1}{2^{n+1}} \prod_{j=0}^n (i x p + {a x^4} + {b x^2} -
                    q - 2n + 2j) \ .
\end{equation}
The Poisson bracket has a form
\[
\{ H_6(p,x), I_n(p,x) \}\ =\ 2p \frac{dI_n(p,x)}{dx} - \frac{dV_6(x)}{dx} x Q(p,x)\ ,
\]
where $Q(p,x)$ is a polynomial in $p, x$ of the degree $n$ in $p$. It is evident that
the Poisson bracket vanishes for trajectories given by
\begin{equation}
\label{st}
 p=0\ , \ \frac{dV_6(x)}{dx}=0\ .
\end{equation}
Therefore, the special trajectories correspond to a system at rest standing at (un)stable
points of equilibrium. Hence, $I_n(p,x)$ is a constant for these trajectories; this is a classical $\pi-$integral. Of course, the values of $H_6, I_n(p,x)$ for these trajectories can be found explicitly. In particular, for the trajectory
$p=x=0$, the energy $H_6 (0,0) = 0$ and the $\pi-$integral $I_n(0,0) = (-)^{n+1} \prod_{j=0}^n (\frac{q}{2} + n - j)$ at $q=1$ and $I_n(0,0) = 0$ at $q=0$.

Similar analysis can be carried out for all other QES problems (\ref{qes}), (\ref{vn}), \ref{Hz}). Always the special trajectories correspond to a particle at rest at the extreme of the potential (\ref{st}).

There exists a class of exactly-solvable multidimensional quantum systems with rational and trigonometric potentials (the Calogero-Sutherland models). Each system is associated to a Lie algebra $g$ of rank $N$, with root space $\De$. In the case of a rational potential the Hamiltonian in the Cartesian coordinates has the form,
\begin{align}
\label{HR}
 {\cal H}^{(r)} = \frac{1}{2}\sum_{k=1}^{N}
 \left[-\frac{\pa^{2}}{\pa x_{k}^{2}}+ \om^2 x^2_k \right]\ +
 \frac{1}{2}\sum_{\alpha\in R_{+}}
 \nu_{|\alpha|}(\nu_{|\al|}-1)\frac{|\,\al|^{\,2}}{(\al\cdot x)^2}\ ,
\end{align}
where $R_+ \in \De$ is a set of positive roots in the root space $\De$, $x$ is a position vector and $\nu_{|\al|}$ are coupling constants (parameters) which depend on the root length. The Hamiltonian (\ref{HR}) is invariant with respect to the Weyl (Coxeter) group transformation, which is the discrete symmetry group of the corresponding root space.
If some roots are of the same length, then $\nu_{|\al|}$ have to be equal, if all roots are of the same length like for $A_n$, then all $\nu_{|\al|}=\nu$. The configuration space for (\ref{HR}) is the Weyl chamber. The ground state wave function is written explicitly,
\begin{align}
\label{psi0r}
  \Psi_{0}^{(r)} (x) \ =\ \prod_{\al\in R_+}
  \left|(\alpha\cdot x)\right|^{\nu_{|\al|}}e^{-\om x^2/2}\ ,
\end{align}
where $\sum x^2 = t_2$ is the lowest order (quadratic) Weyl polynomial invariant.

In the case of trigonometric potential the Hamiltonian in the Cartesian coordinates has the form,
\begin{align}
\label{HES}
 {\cal H}^{(t)}\ =\ \frac{1}{2}\sum_{k=1}^{N}
 \left[-\frac{\pa^{2}}{\pa y_{k}^{2}}\right]\ +
 \frac{\beta^2}{8}\sum_{\alpha\in R_{+}}
 \mu_{|\al|}(\mu_{|\al|}-1)\frac{|\,\al|^{\,2}}{\sin^2 \frac{\beta}{2} (\alpha\cdot y)}
\end{align}
where $R_+ \in \De$ is a set of positive roots in the root space $\De$, $y$ is a position vector and $\mu_{|\al|}$ are coupling constants depending on the root length. The Hamiltonian (\ref{HES}) is invariant with respect to the affine Weyl (Coxeter) group transformation, which is the discrete symmetry group of the corresponding root space plus translations. For roots of the same length the coupling constants $\mu_{|\al|}$ are equal.
The configuration space for (\ref{HES}) is the Weyl alcove. The ground state wave function is written explicitly,
\begin{align}
\label{psi0t}
  \Psi_{0}^{(t)} (y) \ =\ \prod_{\al\in R_+}
  \left|\sin \frac{\beta}{2} (\al , y)\right|^{\mu_{|\al|}}\  .
\end{align}

Extensive studies [6]-[15] had led to a conclusion that the gauge-rotated Hamiltonian     $$(\Psi_{0}^{(r,t)})^{-1}\,({\cal H}^{(r,t)}-E_0^{(r,t)}) \Psi_{0}^{(r,t)}\ =\ h^{(r,t)}\ ,$$
in the space of orbits of (affine) Weyl (Coxeter) group becomes an algebraic operator -- a differential operator with polynomial coefficients. Here $E_0^{(r,t)}$ is the ground state energy of rational (trigonometric) model. It implies that a change of variables should be done from the Cartesian coordinates to the {\it Weyl (Coxeter) polynomial invariants},
\begin{equation}
\label{Pol_Inv}
 t_{a}^{(\Om)}(x) = \sum_{\al\in\Om} (\al \cdot x)^{a}\ ,
\end{equation}
where $a$'s are the {\em degrees} of the Weyl (Coxeter) group and $\Om$
is an orbit, for the case of a rational potential.
In the case of a trigonometric potential it has to be taken as new variables the {\it Weyl (Coxeter) trigonometric  invariants},
\begin{equation}
\label{Trig_Inv}
 \ta_{a}(y) = \sum_{\om \in\Om_a} e^{i \beta (\om , y)}\ ,
\end{equation}
where $\Om_a$ is the orbit generated by fundamental weight $w_a$
\footnote{For a given root space $\De$ and a fixed $\beta$, there thus exist $N$ independent trigonometric Weyl invariants $\tau_a$ generated by $N$ fundamental weights $w_a$.}. Furthermore, it was shown that any such algebraic operator $h^{(r,t)}$ has infinitely-many finite-dimensional invariant subspaces in a form of the "triangular" linear space of polynomials
\begin{equation}
\label{InvSpace}
 {\cal P}^{(d)}_{n, \vec f} \ = \ \langle t_1^{p_1}
t_2^{p_2} \ldots t_d^{p_d} | 0 \leq \underbrace{f_1 p_1 + f_2 p_2 +\ldots +
f_d p_d}_{\mbox{grading}} \leq n \rangle\
\end{equation}
where the ``grades" $f$'s are positive integer numbers which are ordered in
such a way that they are non-decreasing with the growth of $i$. If a notion of the characteristic vector is introduced
\[
 \vec f = (f_1, f_2, \ldots f_d)\ ,
\]
it was shown that for a given root space $\De$, the vector $\vec f$ is either the highest root or, the Weyl vector or, Weyl co-vector with the only exception for $E_8$ rational case where $\vec f$ is special (for a discussion see \cite{BTLG}). For any space (\ref{InvSpace}) one can indicate an infinite-dimensional algebra of differential operators for which ${\cal P}^{(d)}_{n, \vec f}$ is a common invariant subspace. Such an algebra is called {\it hidden}. It is evident that the algebraic operator $h^{(r,t)}$ is an element of the hidden algebra. For all studied root spaces this algebra is finitely-generated. It is found that among generating elements of the hidden algebra, the most symmetric generator (the Euler-Cartan operator) is always present
\begin{equation}
\label{cartan}
   {\cal J}^0_n\ =\ \sum_{i=1}^{d} f_i t_i \frac{\pa}{\pa t_i} - n\, ,
\end{equation}
which has zero grading and plays a role of constant acting as identity operator.
It defines the highest weight vector. This generator allows us to construct the $\pi$-integral of zero grading
\begin{equation}
\label{pics}
    i_n(t)\ =\ \prod_{j=0}^n ({\cal J}^0_n + j)
\end{equation}
such that
\begin{equation}
\label{hid}
  [h^{(r,t)}, i_n(t)]: {\cal P}^{(d)}_{n, \vec f} \ \mapsto \ 0\ .
\end{equation}
This $\pi$-integral has no finite-dimensional invariant subspaces of the dimension higher
than $\dim({\cal P}^{(d)}_{n, \vec f})$. Making a gauge rotation of (\ref{pics}) with $\Psi_0^{(r,t)}$ (see (\ref{psi0r}), (\ref{psi0t}), respectively) and changing variables $t, \ta$ (see (\ref{Pol_Inv}), (\ref{Trig_Inv}), respectively) back to the Cartesian coordinates we arrive at the quantum $\pi$-integral ${\cal I}_n^{(r,t)} (x)$. It is a differential operator of the $(n+1)$th order. Under such a transformation the triangular space of polynomials ${\cal P}^{(d)}_{n, \vec f}$ becomes the space
\[
     {\cal V}^{(d)}_{n, \vec f}\ =\  \Psi_0^{(r,t)} {\cal P}^{(d)}_{n, \vec f} \ .
\]
Needless to say that
the Hamiltonian ${\cal H}^{(r,t)}(x)$ commutes with ${\cal I}_n^{(r,t)}(x)$,
\[
     [{\cal H}^{(r,t)}(x)\ ,\ {\cal I}_n^{(r,t)}(x)]: {\cal V}^{(d)}_{n, \vec f}\ \mapsto \ 0\ .
\]
Replacing the operator of momentum by the classical momentum (\ref{quant}) in
${\cal H}^{(r,t)}$ and ${\cal I}^{(r,t)}_n(x)$ we get a classical Hamiltonian $H^{(r,t)}(p,x)$ and a classical analogue of the $\pi-$integral $I^{(r,t)}_n(p,x)$.
It can be easily checked that the Poisson bracket $\{ H^{(r,t)}, I^{(r,t)}_n \}$
vanishes for trajectories which correspond to the particle at extremes of the potential
at zero momentum similar to (\ref{st})).

In this paper we introduced a notion of particular integrability, both quantum and classical.
In quantum mechanics particular integrability implies the existence of a certain
number of common eigenfunctions of the Hamiltonian and a particular integral. In classical mechanics it realizes the idea of the existence of a number of (special) trajectories which are characterized by a particular constant of motion. Hence, in such a way the property of particular integrability is preserved under the quantization, although a number of common eigenfunctions does not seem related with a number of special trajectories.
We find an explicit form for particular integral for all one-dimensional (quasi)-exactly-solvable problems as well as for Calogero-Sutherland models for all root spaces. In both cases the particular integral looks especially easy in action on the space of polynomials parameterized by symmetric coordinates (it coincides to the space of orbits in the case of Weyl (Coxeter) symmetry). Its common eigenfunctions with the gauge-rotated Hamiltonian written in symmetric coordinates are given by finite-degree polynomials in those coordinates. Special trajectories correspond to standing particle at an extreme of the potential. The situation gets different for the case of the Magnetic Hamiltonians. For example, for two charges on the plane subject to a constant magnetic field the special trajectories appear as "well organized" among, in general, chaotic motion: they are either simply closed, or concentric, or parallel corresponding to a free motion \cite{ET-c}. For each of these trajectories a certain $\pi-$particular constants of motion can be indicated. In the quantum case the common eigenfunctions of the Hamiltonian and $\pi-$particular integral(s) have the form of polynomial multiplied by some factor \cite{ET-q}.

\textit{\small Acknowledgements}.  The author is grateful to S.P.~Novikov and P.~Winternitz for the interest to work and valuable discussions.
This research is supported in part by DGAPA grant IN109512 (Mexico).  The author also thanks the University Program FENOMEC (UNAM, Mexico) for partial support.

\end{document}